\documentclass[]{pasj01}
\usepackage{braket}
\usepackage{color}

\begin{document} 
\Received{}
\Accepted{}

\title{\textcolor{black}{Large} velocity dispersion of molecular gas in bars of strongly barred galaxies, NGC 1300 and NGC 5383}

\author{Fumiya \textsc{Maeda}\altaffilmark{1}}
\altaffiltext{1}{Department of Astronomy, Kyoto University, Kitashirakawa-Oiwake-Cho,
 Sakyo-ku, Kyoto, 606-8502, Japan}
\email{fmaeda@kusastro.kyoto-u.ac.jp}

\author{Kouji \textsc{Ohta}\altaffilmark{1}}

\author{Yusuke \textsc{Fujimoto}\altaffilmark{2}}
\altaffiltext{2}{Research School of Astronomy \& Astrophysics, Australian National University, Canberra, ACT, Australia 2611}

\author{Asao \textsc{Habe}\altaffilmark{3}}
\altaffiltext{3}{Graduate School of Science, Hokkaido University, Kita 10 Nishi 8, Kita-ku, Sapporo 060-0810, Japan}

\author{Junichi \textsc{Baba}\altaffilmark{4}}
\altaffiltext{4}{National Astronomical Observatory of Japan (NAOJ), 2-21-1 Osawa, Mitaka, Tokyo 181-8588}


\maketitle

\begin{abstract}
We carried out  $\sf ^{12}$CO($\sf J = 1 -0$) observations toward bar and arm regions of strongly barred galaxies,
NGC 1300 and NGC 5383, with the Nobeyama 45-m radio telescope (beamsize of $\sf 1-2$ kpc in the galaxies).
The aim of the observations is to qualitatively examine a new scenario for the suppression of star formation
in bars based on recent high-resolution numerical simulations:
higher speed collisions between molecular clouds in the bar region compared with the arm region
suppress the massive star formation.
CO emissions were detected from all the regions,
indicating the presence of the molecular gases in the
strong bars without associating clear HII regions.
In both galaxies,
the velocity width of the CO line profile tends to be larger in the bar region than in the arm region,
which is qualitatively consistent with the new scenario.
\end{abstract}

\section{Introduction}
Galaxies have been evolving by converting gas into stars. 
In the present-day universe,
stars form mostly in spiral arms as well as in the central region of a galaxy.
In spiral arms, stars form in giant molecular clouds (GMCs);
their masses are $\sim 10^{5-6}~M_\odot$ and sizes are $\sim 10-100$ pc (e.g., \cite{Solomon87}).
GMCs coexist with dust lanes,
and HII regions are associated with the dust lanes (e.g., M51; \cite{Schinnerer13,Colombo14}).
This is usually interpreted in terms of the spiral density wave;
the gases go into the spiral density wave, 
then dense shocked gas regions form in the spiral arms which result in the formation of GMCs.
In the GMCs, stars form and the massive stars produce HII regions.

However, star formation in bar regions of typical strongly barred galaxies
is different from that in spiral arms.
Although there are remarkable dust lanes along the stellar bar,
prominent HII regions are often not seen. 
The presence of the dust lane implies the existence of molecular clouds.
Nevertheless, (massive) star formation is notably absent.

What prevents star formation in bar regions? 
This question is the long-standing problem,
and several explanations have been proposed.
\citet{Tubbs82} suggested the molecular clouds
may be destroyed by shock due to the high velocity of the gases relative to the bar structure.
\citet{Athanassoula92} argued that strong shear motion
in a bar region prevents molecular cloud formation. 
Combination of the shock and the shear
may be the cause for the low star formation activity \citep{Reynaud98}.
\citet{Sorai12} suggested that molecular gas in bar regions
is gravitationally unbound.
However,
the cause for the suppression of star formation in bar regions has not been clear.

Recent hydrodynamic numerical simulations with 
a very high spatial resolution of $\sim$ 1.5 pc \citep{Nimori13,Fujimoto14a}
revealed that GMC-like gas clouds exist both in arms and bars in barred galaxies,
and the clouds collide with each other, which triggers star formation.
The collision velocity between the clouds is
$10 \sim 40 \rm ~km~s^{-1}$ in the arm and disk regions,
while in the bar region a larger ($> 50 \rm~km~s^{-1}$) collision velocity is seen. 
\citet{Fujimoto14b} proposed a new scenario to prevent massive star formation in the bar regions:
The cloud collision forms cloud cores both in the arm region and the bar region,
but the high-speed collision of the clouds in the bar region
shortens the gas accretion phase of the cloud cores formed,
leading to suppression of core growth and massive star formation
 \citep{Takahira14,Takahira17}. 
According to these results, 
observational keys to understanding the star formation  suppression in the bar
regions are 
(i) identifying an individual GMC-like molecular cloud in a bar region,
(ii) deriving the amount of velocity dispersion among the clouds, 
and (iii) comparing them with the characteristics of molecular gases in the arm regions.

CO observations towards nearby barred galaxies have been made:
e.g., M83 \citep[etc.]{Handa90,Lord91,Rand99,Lundgren04,Sakamoto04,Muraoka07,Muraoka09,Hirota14},
NGC 1097 \citep[etc.]{Gerin88,Hsieh11},
NGC 1300 \citep{Regan99},
NGC 1365 \citep{Sandqvist95,Sandqvist99},
NGC 1530 \citep{Regan99,Reynaud97,Reynaud99},
NGC 2903 \citep{Regan99,Muraoka16},
NGC 3627 \citep{Regan99}, 
NGC 4303 \citep{Momose10},
NGC 4314, NGC 5135 \citep{Regan99},
NGC 5383 \citep{Ohta86,Regan99,Sheth00},
NGC 7479 \citep{Sempere95,Laine99},
and other barred galaxies in CO surveys \citep{Regan01,Kuno07,Bolatto17}.
However, almost all of these observations were
made with large beamsizes even with the interferometer,
corresponding to a physical size of typically 1 kpc or more.
Further, many of the observations focused on the central regions
and the arm regions in the barred galaxies,
and high resolution and high sensitivity observations of the bar regions
have been very much limited. 
Unfortunately, these galaxies previously observed are not necessarily
suitable targets to study the cause for the suppression of star formation
activity in the bar regions, because many of them have an intermediate-type bar
(classified as SAB or even SA) and do have star forming regions
associated with the bar or bar-like region.

In order to understand the cause
for the suppression of the star formation in the bar regions,
CO observations of strongly barred galaxies 
that do not show massive star formation in the bar regions are important,
since the effect of high-speed collisions is expected to be clearly seen in such bars.
Atacama Large Millimeter/submillimeter Array (ALMA) and
NOrthern Extended Millimeter Array (NOEMA)
enable us to examine the velocity distribution
among individual GMC in nearby galaxies
thanks to the high angular resolution and sensitivity.
Prototype strongly barred galaxies NGC 1300 and NGC 5383
are very suitable laboratories to examine the properties
of the molecular clouds to test the new scenario described above.
In both galaxies, remarkable dust lane 
is seen in the bar region without prominent HII regions,
while in the arm region HII regions are associated with dust lane.
Since these galaxies are comparatively near to us (see table \ref{tab:first}),
we can achieve a high spatial resolution of $50 - 70$ pc, corresponding
to the typical GMC size with the ALMA/NOEMA.
However, 
CO emission line has not been detected
towards the bar regions of both galaxies
(e.g., \cite{Ohta86,Regan99}).

\begin{figure}
 \begin{center}
  \includegraphics[width=8.5cm]{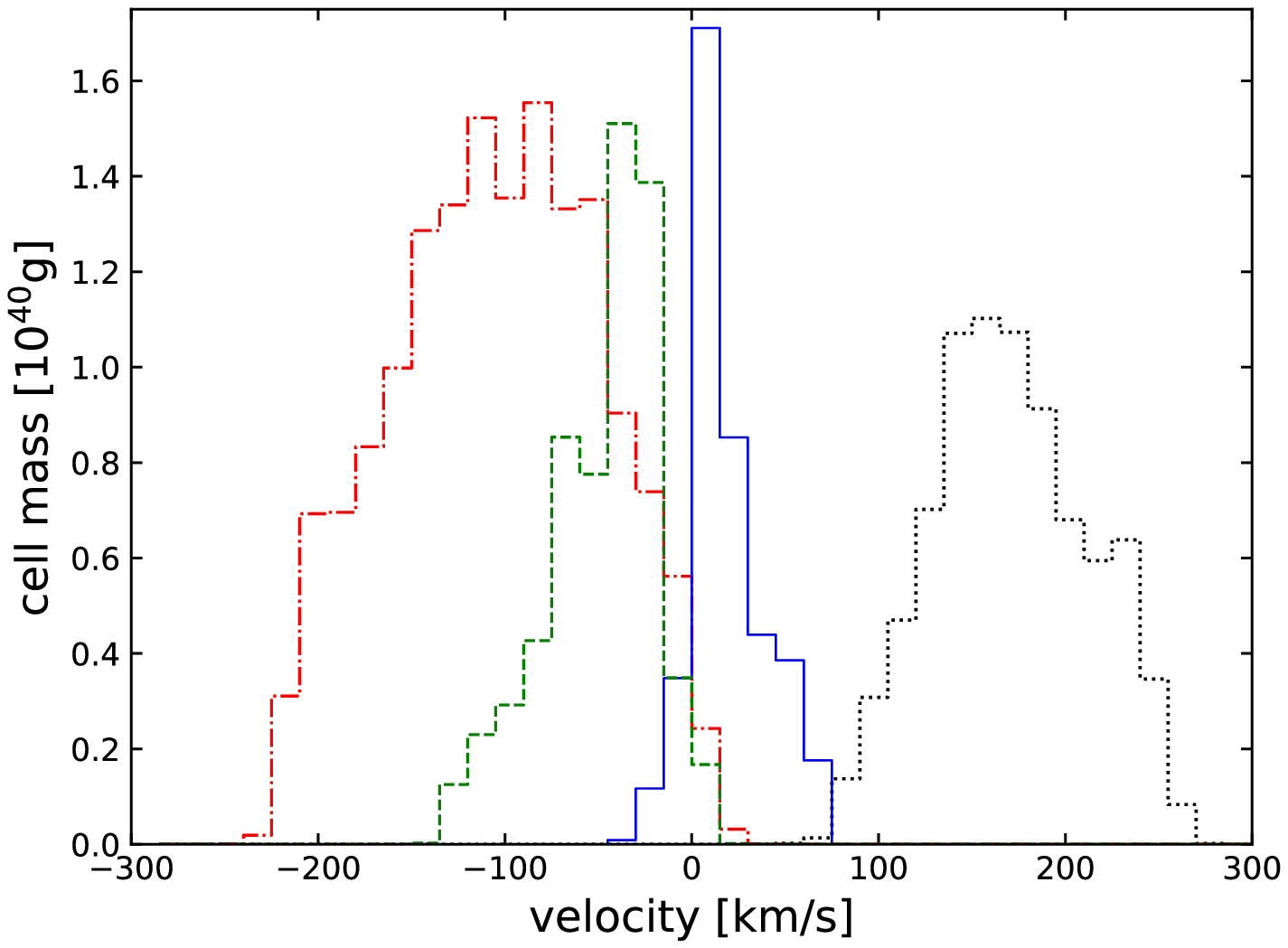}
 \end{center}
\caption{Line profiles of the gases in the bar regions (black solid line and red dash-dotted line) and the arm regions (blue dotted line and green dashed line) in the simulation by \citet{Fujimoto14a} and \citet{Fujimoto14b}.
The bar (arm) regions are selected from both sides with respect to the center of the galaxy.
Each region is a circular region with a diameter of 1.0 kpc and the viewing angle is 35$^\circ$ which is the same as that of NGC1300. 
The width of each  bin is 15 $\sf km~s^{-1}$.
The offset of the velocity is just reflection of the velocity field (rotation curve).
}\label{fig:fujimoto}
\end{figure}

In this paper, we report results of $^{12}{\rm CO} (J = 1-0)$ emission
line observations of NGC 1300 and NGC 5383 with the 45-m telescope
of Nobeyama Radio Observatory (NRO) \footnote{Nobeyama Radio Observatory (NRO) is a branch of the National Astronomical Observatory of Japan, National Institutes of Natural Sciences.}.
The purposes of the observations are 
(1) to detect molecular gases in the bar regions and arm regions
in NGC 1300 and NGC 5383 and to derive masses of the molecular gases,
to make further investigations with ALMA/NOEMA in the future,
and (2) to derive velocity widths of CO emissions in a beamsize ($\sim 1 -2$ kpc) of the 45-m telescope,
and compare the velocity widths in the bar regions and arm regions.
The purpose (2) intends to examine the scenario described by \citet{Fujimoto14b};
Figure \ref{fig:fujimoto} illustrates line profiles of
the gases ($n \geq 100~\rm cm^{-3}$)
in the bar and arm regions in the simulation.
The bar (arm) regions are selected from both sides
with respect to the center of the galaxy.
Each region is taken to represent the beamsize of NRO 45-m telescope,
i.e., a circular region with a diameter of 1.0 kpc,
and we set the viewing angle of $35^\circ$
which is the same as the inclination of NGC1300.
Since the relative velocity among the clouds in the bar regions is larger than that in the arm regions, 
the profiles in the bar regions (black and red) are much wider (Full Width at Zero Intensity, FWZI $\sim 250~\rm km~s^{-1}$) 
than those in the arm regions (blue and green) ($\sim 130~\rm km~s^{-1}$). 
Although the galaxy in this simulation was modeled based not on a strongly barred galaxy but an intermediate-type one (M83),
we investigate whether the same tendency is qualitatively seen in observed profiles or not.

In section \ref{sec:obs}, we describe the details of the CO observations
of NGC 1300 and NGC 5383 and show results.
In section \ref{sec:discussion},
we compare resulting
velocity widths in the bar regions with those in the arm regions.
Finally, we give a summary in section \ref{sec:summary}.
Table \ref{tab:first} summarizes parameters of
both galaxies adopted throughout this paper.

\begin{table*}[!t]
  \tbl{Properties of NGC 1300 and NGC 5383}{%
  \begin{tabular}{cccccccc}
      \hline
              & \multicolumn{2}{c}{Coordinates of center}      & Morphology  & Inclination & Systemic velocity$^{a)}$  & Distance$^{b)}$  & Liner scale \\
              & R.A.(J2000)            & Dec.(J2000)           &              & (degree)    & (km s$^{-1}$)             & (Mpc)            & (pc arcsec$^{-1}$)\\
      \hline \hline
      NGC 1300 & $\rm 03^h19^m41^s.11$  & $\rm -19^\circ24^\prime28^{\prime\prime}.4$$^{1)}$ & SB(s)bc$^{2)}$ & $35$$^{1)}$    & $1557$$^{1)}$             & $20.5$           & $99.2$\\
      NGC 5383 & $\rm 13^h57^m04^s.81$  & $\rm +41^\circ50^\prime46^{\prime\prime}.5$$^{3)}$ & SB(s)b$^{2)}$  & $40$$^{4)}$    & $2260$$^{2)}$             & $31.9$           & $154.4$\\
      \hline
    \end{tabular}}\label{tab:first}
\begin{tabnote}
$^{a)}$ The local standard of rest (LSR). \\
$^{b)}$ We adopted the Hubble constant of $73~{\rm km~s^{-1}~Mpc^{-1}}$. \\
References: $^{1)}$ \citet{Lindblad97},$^{2)}$ \citet{SandageandTammann},
$^{3)}$ \citet{Sheth00},
$^{4)}$ \citet{vanderKruit}
\end{tabnote}
\end{table*}

\begin{figure*}
 \begin{center}
  \includegraphics[width=15cm]{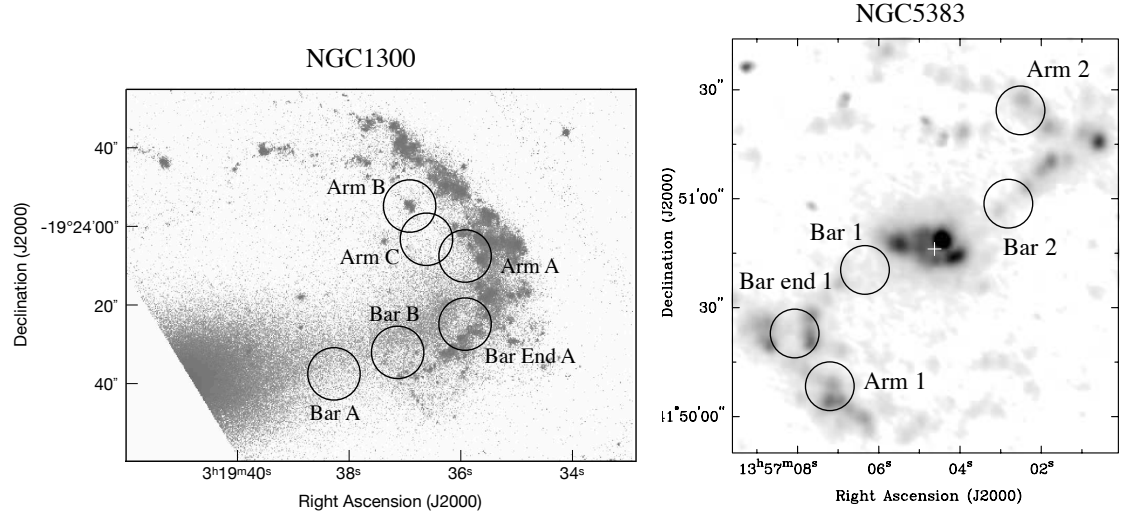} 
 \end{center}
\caption{NGC 1300 taken with Hubble Space Telescope (F658N : H$\sf \alpha$ emission + continuum) (left). Continuum subtracted H$\sf \alpha$ image of NGC5383 (edited from Figure 1 in \citet{Sheth00} by courtesy of K. Sheth).
Observed positions are superimposed. Circles show the beamsize of 13.5$^{\prime\prime}$ of the NRO 45-m telescope.}\label{fig:NGC1300_5383}
\end{figure*}

\section{Observations and Results}\label{sec:obs}
\subsection{Observations with the Nobeyama 45-m telescope}
We carried out $^{12}{\rm CO} (J = 1-0)$ line (rest frequency $=$ 115.271204 GHz)
observations of NGC 1300 
on 2016 January 31, February 1, April 24 - 25, and 2017 February 14 - 15
and NGC 5383 on 2017 February 14 - 18
with the Nobeyama 45-m telescope.
The observed positions are shown with circles
in figure \ref{fig:NGC1300_5383} superimposed on H$\alpha$ images.
We observed six regions in NGC 1300.
Bar A and Bar B are in the bar region 
where the dust lane is most clearly seen and there are no HII regions associated with it.
Arm A and Arm B are in the arm region where clear dust lane associated with HII regions is seen.
Arm C is also in the arm region but with no HII regions.
We additionally observed a bar end region (Bar End A) where HII regions are seen.
In NGC 5383, we observed five regions. 
Bar 1 and Bar 2 include dust lanes with no clear HII regions,
and Arm 1 and Arm 2 are on the dust lane with HII regions.
We also observed Bar End 1.
The coordinates of the observed regions are listed in table \ref{tab:obsresult}.

We used one-beam (TZ1) of the two-beam dual-polarization (H, V) sideband-separating SIS receiver
(TZ receiver: \cite{Nakajima13}).
The half-power beam width (HPBW) was $\sim 13.5^{\prime\prime}$
which corresponds to 1.34 kpc and 2.08 kpc at the distance of NGC 1300 and NGC 5383, respectively.
The backend was an FX-type correlator system, SAM45,
which consisted of 16 arrays with 4096 spectral channels each. 
Four arrays (A1 - A4) observed polarization V,
and the others (A5 - A16) observed polarization H.
The bandwidth and frequency resolution was 2.0 GHz and 488.28 kHz, respectively,
which correspond to 5217 $\rm km~s^{-1}$ and 1.27 $\rm km~s^{-1}$  at 115 GHz.
We employed the position-switching mode with an on-source integration time of 20 sec per scan.
The telescope pointing was checked every $\sim$ 50 scans ($\sim$ 50 min) by observing SiO masers
near the targets.
The typical pointing error was  $3^{\prime \prime} - 5^{\prime \prime}$ 
and $\sim 2^{\prime \prime}$ for NGC 1300 and NGC 5383, respectively. 
The line intensity was calibrated by the chopper wheel method.
The $T_{\rm sys}$ was 
$350 - 600$ K and $200 - 300$ K for NGC 1300 and NGC 5383, respectively. 
The typical pointing error and $T_{\rm sys}$ for NGC 1300
were large due to the low elevation of $20^\circ - 30^\circ$.

\subsection{Data reduction}
The observed data were analyzed using the NRO reduction software, NEWSTAR.
We flagged scans taken under a wind velocity larger than 8 $\rm m~s^{-1}$ and with poor baselines
by inspecting each spectrum by eye.
We examined three flagging criteria to check the robustness of our results.
The percentage of flagged data and effective on-source integration time
after flagging for each region were $10-20$ \% and $0.25-2$ hr, respectively.
We combined both polarizations;
we examined the various combinations of the arrays for V and H
to see the robustness of the results.
We subtracted a baseline (second-order polynomial function with small curvature) which was determined
by fitting to the combined spectrum
except for the signal frequency range.
After the baseline subtraction, we smoothed the spectrum by binning to 10 $\rm km~s^{-1}$.
The image rejection ratios (side band ratio) in the adopted frequency range were larger than 10 dB,
and hence no correction was made.
We converted the antenna temperature ($T_{A}^\ast$) 
into the main beam brightness temperature ($T_{\rm mb}$) 
using the main beam efficiency of $\eta_{\rm mb} = 0.42$, where $T_{\rm mb} = T_{A}^\ast / \eta_{\rm mb}$.

\begin{table*}[htpb]
  \tbl{Observation results}{%
  \begin{tabular}{clrrrrrrrrr}
  \hline
\multicolumn{2}{c}{Region}    & \multicolumn{2}{c}{Coordinates of center}   & \multicolumn{1}{c}{$t_{\rm on}^{\rm eff}$$^{\ast)}$} & \multicolumn{1}{c}{$T_{\rm rms}$$^{\dagger)}$} &\multicolumn{1}{c}{$T_{\rm peak}$}         & \multicolumn{1}{c}{$I_{\rm CO}$}& \multicolumn{1}{c}{$\Sigma_{\rm mol}$}        & \multicolumn{1}{c}{$M_{\rm mol}$}    \\
         &           &\multicolumn{1}{c}{R.A.(J2000)}            & \multicolumn{1}{c}{Dec.(J2000)} & \multicolumn{1}{c}{(hr)}                             & \multicolumn{1}{c}{(mK)}                       &  \multicolumn{1}{c}{(mK)}  &  \multicolumn{1}{c}{($\rm K~km~s^{-1}$)}     & \multicolumn{1}{c}{($M_\odot~{\rm pc^{-2}}$)} & \multicolumn{1}{c}{($10^7~M_\odot$)}\\
  \hline
  \hline
NGC 1300 & Bar A     & $\rm  3^h19^m38^s.26$ & $\rm -19^\circ24^\prime37^{\prime\prime}.6$ & $2.15$ & $8.1$  & $47.5$  & $ 2.8\pm0.3$ & $10.2 \pm 1.1$ & $1.7 \pm 0.2$ \\
         & Bar B     & $\rm  3^h19^m37^s.10$ & $\rm -19^\circ24^\prime31^{\prime\prime}.9$ & $1.11$ & $9.5$  & $110.8$ & $ 4.7\pm0.3$ & $17.0 \pm 1.1$ & $2.9 \pm 0.2$ \\
         & Arm A     & $\rm  3^h19^m35^s.90$ & $\rm -19^\circ24^\prime07^{\prime\prime}.5$ & $0.47$ & $12.9$ & $120.7$ & $ 4.9\pm0.4$ & $17.7 \pm 1.3$ & $3.0 \pm 0.2$ \\
         & Arm B     & $\rm  3^h19^m36^s.90$ & $\rm -19^\circ23^\prime54^{\prime\prime}.7$ & $1.68$ & $11.8$ & $69.1$  & $ 3.0\pm0.4$ & $10.7 \pm 1.4$ & $1.8 \pm 0.2$ \\
         & Arm C     & $\rm  3^h19^m36^s.60$ & $\rm -19^\circ24^\prime03^{\prime\prime}.3$ & $0.39$ & $15.4$ & $89.0$  & $ 3.2\pm0.4$ & $11.4 \pm 1.6$ & $2.0 \pm 0.2$ \\
         & Bar End A & $\rm  3^h19^m35^s.90$ & $\rm -19^\circ24^\prime24^{\prime\prime}.7$ & $0.48$ & $15.9$ & $113.0$ & $ 6.3\pm0.5$ & $22.5 \pm 1.7$ & $3.9 \pm 0.3$ \\
\hline
NGC 5383 & Bar 1     & $\rm 13^h57^m06^s.40$ & $\rm +41^\circ50^\prime40^{\prime\prime}.9$ & $0.77$ & $ 8.4$ & $135.6$ & $13.5\pm0.4$ & $44.9 \pm 1.4$ & $20.0\pm 0.6$  \\
         & Bar 2     & $\rm 13^h57^m03^s.10$ & $\rm +41^\circ50^\prime59^{\prime\prime}.6$ & $0.27$ & $12.2$ & $118.3$ & $ 8.6\pm0.5$ & $28.7 \pm 1.6$ & $12.8\pm 0.7$  \\
         & Arm 1     & $\rm 13^h57^m07^s.20$ & $\rm +41^\circ50^\prime08^{\prime\prime}.0$ & $2.00$ & $ 5.8$ & $ 43.4$ & $ 2.9\pm0.2$ & $ 9.8 \pm 0.8$ & $ 4.3\pm 0.3$  \\
         & Arm 2     & $\rm 13^h57^m02^s.70$ & $\rm +41^\circ51^\prime24^{\prime\prime}.3$ & $1.88$ & $ 4.9$ & $ 46.3$ & $ 2.3\pm0.2$ & $ 7.8 \pm 0.6$ & $ 3.5\pm 0.2$  \\
         & Bar End 1 & $\rm 13^h57^m08^s.10$ & $\rm +41^\circ50^\prime23^{\prime\prime}.3$ & $0.86$ & $ 8.3$ & $ 34.6$ & $ 3.8\pm0.4$ & $12.7 \pm 1.2$ & $ 5.6\pm 0.5$  \\
  \hline
    \end{tabular}}\label{tab:obsresult}
\begin{tabnote}
$^{\ast)}$ The effective on-source integration time.\\
$^{\dagger)}$ 1 $\sigma$ r.m.s at 10 $\rm km~s^{-1}$ bin.\\
\end{tabnote}
\end{table*}

\subsection{Results}
\begin{figure*}[!h]
 \begin{center}
  \includegraphics[width=160mm]{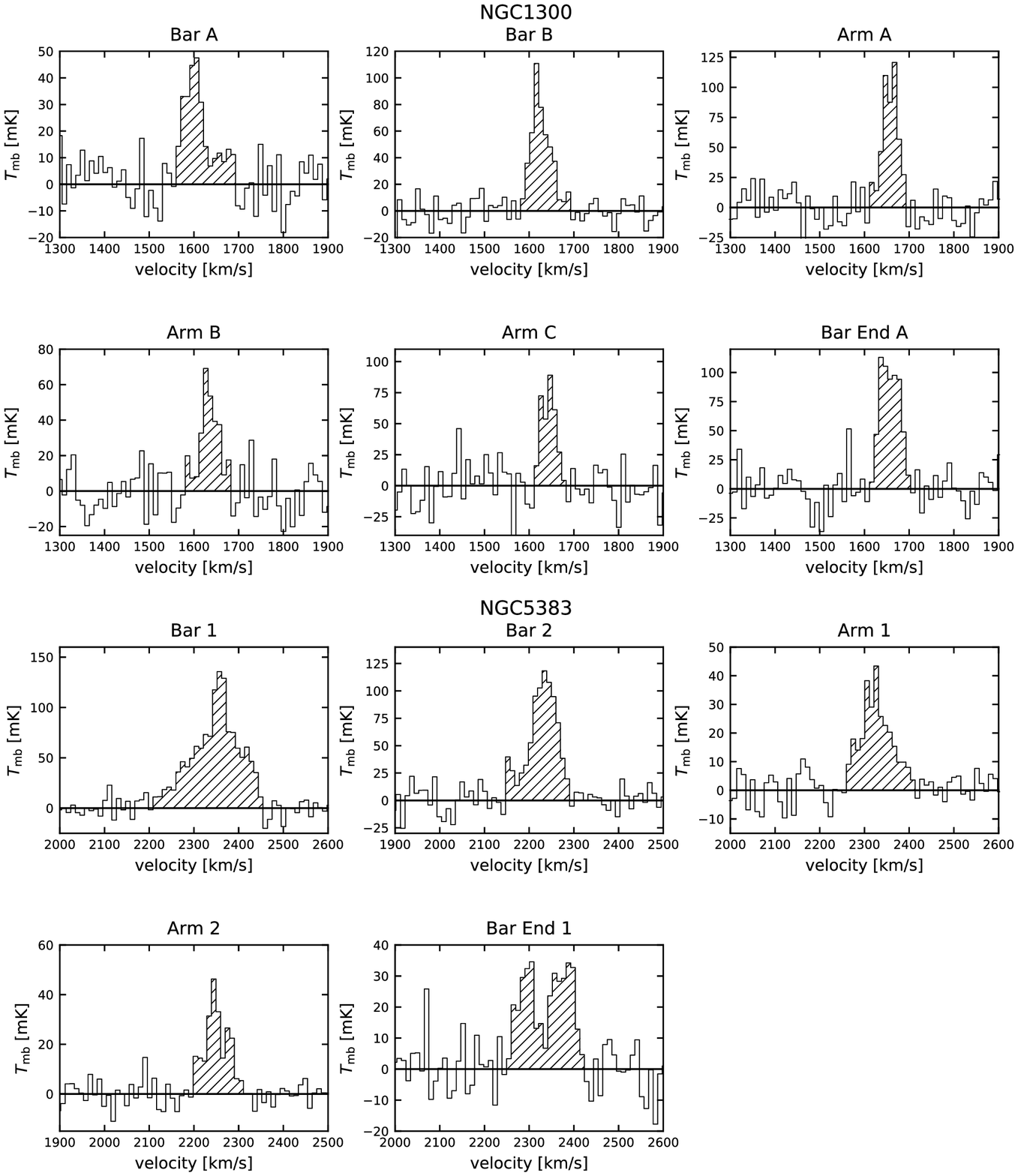} 
 \end{center}
\caption{CO emission profiles in NGC 1300 and NGC 5383 obtained with Nobeyama 45-m
telescope. The observed positions are shown in figure \ref{fig:NGC1300_5383} with black circles.
Velocity refers to LSR velocity. 
The CO emission line is shown as a hatched region.}\label{fig:result}
\end{figure*}

We detected  CO emission lines from all the regions observed.
All profiles of $^{12}{\rm CO}(J=1-0)$ after binning to 10 $\rm km~s^{-1}$ in NGC 1300 and NGC 5383 
are presented in figure \ref{fig:result} and observation results are summarized in table \ref{tab:obsresult}.
Significant emissions were seen in
all the bar regions in NGC 1300 and NGC 5383, indicating that
the molecular gases do exist in the bar
regions\footnote{\citet{Ohta86} made CO observations toward the northwestern bar region
(mostly the same region of Bar 2) of NGC 5383 with the 45-m telescope
but did not detect significant emission.
This is considered to be due to the low sensitivity at that time;
$T_{\rm sys}$ was much higher, the on-source time was short ($\sim$ 10 min.), and only one polarization was used.
Non-detection of CO emission in the bar regions in NGC 1300 and NGC 5383 by \citet{Regan99} and \citet{Sheth00} is also due to the low sensitivity.}.

For the bar regions, most of the spectra show asymmetric shapes similar
to the  results of the simulation (Figure \ref{fig:fujimoto})
and some of them show double peak structure (Bar A and Bar End 1).
The CO line widths tend to be wider in the bar regions than
in the arm regions as seen in the simulation.
More details are  described in section \ref{sec:discussion}.

We derived velocity-integrated intensities ($I_{\rm CO}$) from the emission shown as hatched region in figure \ref{fig:result},
and then derived the surface density of the molecular gas by
\begin{equation}\label{eq:Sigmamol}
\Sigma_{\rm mol} = \alpha_{\rm CO} I_{\rm CO} \cos i
\end{equation}
where $\alpha_{\rm CO}$ is CO-to-H$_2$ conversion factor and 
$\cos i$ shows the effect of inclination of the galaxy.
Here, we adopted a $\alpha_{\rm CO}$ value of $4.36~M_\odot~(\rm K~km~s^{-1}~pc^2)^{-1}$
including a factor of 1.36 to account for the presence of helium
($X_{\rm CO} = 2.0 \times 10^{20} \rm~cm^{-2}~(K~km~s^{-1})^{-1}$).
The molecular gas mass in the beamsize was derived by
\begin{equation}\label{eq:Mmol}
M_{\rm mol} = \alpha_{\rm CO} I_{\rm CO} \pi R^2
\end{equation}
where $R$ is radius of the observed region (HPBW).
The $R$ is 670 pc and 1040 pc in NGC 1300 and NGC 5383, respectively.

Resulting $I_{\rm CO},~\Sigma_{\rm mol},~M_{\rm mol}$ in the observed positions are shown in table \ref{tab:obsresult}.
The molecular gas masses in the beamsize are $(1.7 - 3.9) \times 10^7 M_\odot$ in NGC 1300 and 
$(3.5 - 20.0) \times 10^7 M_\odot$ in NGC 5383.
It is worth noting that the molecular gas mass in the bar regions may have an uncertainty.
Using large velocity gradient analysis, \citet{Sorai12} suggested that molecular gas 
in the bar regions may be gravitationally unbound
and $\alpha_{\rm CO}$ may be $0.5 - 0.8$ times smaller than that 
in the arms in Maffei II.
\citet{Morokuma15} suggested the existence of non-optically thick
components of $^{12}$CO$(1-0)$ in the bar regions may make the  $\alpha_{\rm CO}$ 
by a factor of a few smaller
than in the arm regions in NGC 3627.
Therefore, it is possible that the $\Sigma_{\rm mol}$ and $M_{\rm mol}$
in the bar regions of NGC 1300 and NGC 5383 may be overestimated by a factor of a few.

In NGC5383, the surface densities in the bar regions are
higher than that in the arm regions by a factor of $3 \sim 5$,
whereas those in the bar regions and the  arm regions are comparable in NGC 1300.
The cause of this difference is not clear,
but it is possible that this result shows a difference in the radial distribution of molecular gases.
\citet{Jogee05}, \citet{Sheth05}, and \citet{Kuno07} suggested that a
bar component in barred spiral galaxies drives a gas inflow,
which changes the radial distribution of the gas.
The presence of the active star formation in the central region of NGC 5383 \citep{Sersic73}
may be related to the gas inflow.
It is interesting to investigate the radial distribution of the molecular gases,
but it is beyond our scope in this paper.

\begin{figure*}
 \begin{center}
  \includegraphics[width=145mm]{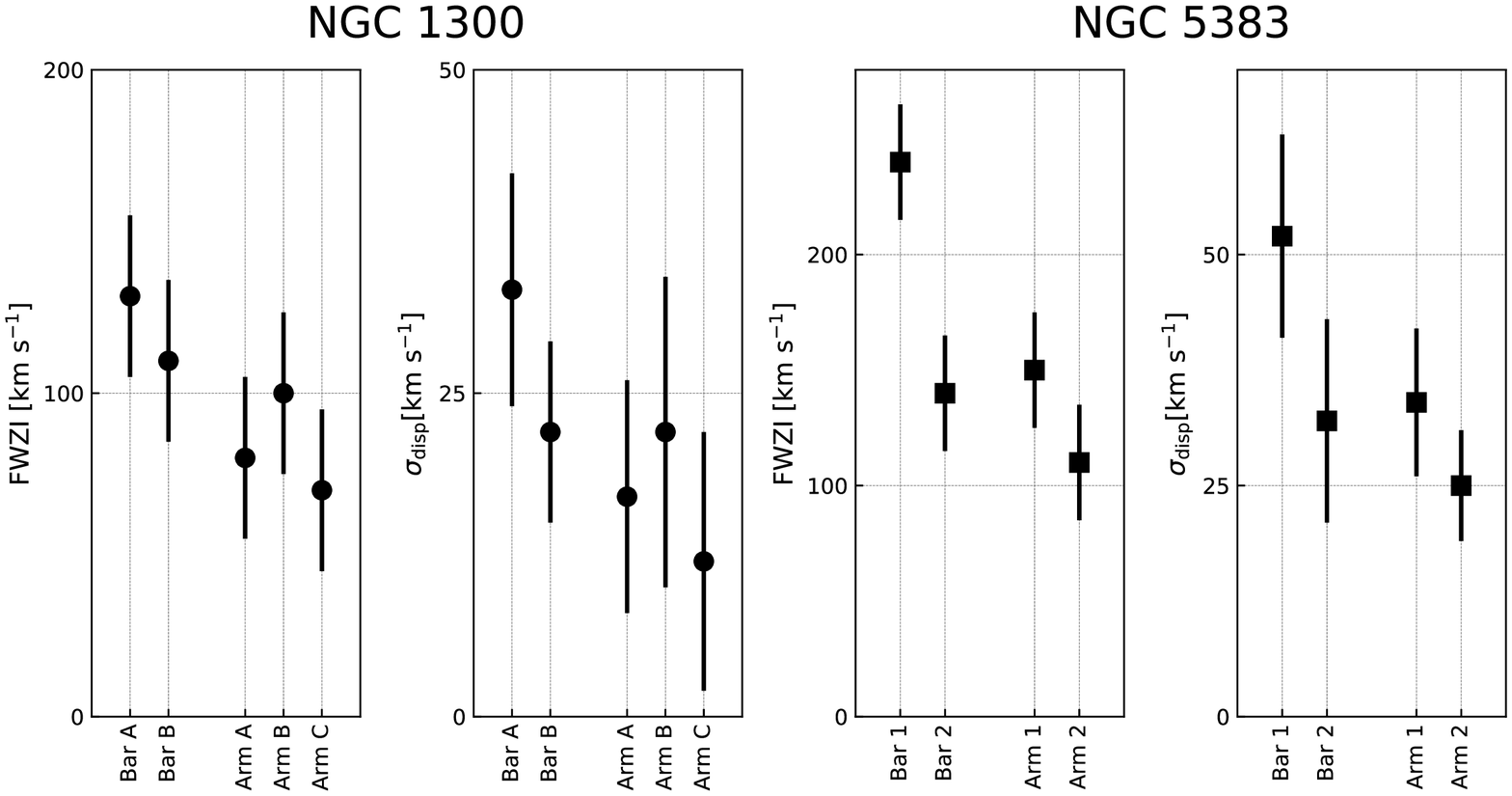} 
 \end{center}
\caption{Velocity widths (Full Width at Zero Intensity, FWZI, and intensity-weighted velocity dispersion, $\sf \sigma_{disp}$)
in the bar and arm regions of NGC 1300 (left panel) and NGC 5383 (right panel).
Error bars shown for the FWZI are $\sf 25~km~s^{-1}$.}
\label{fig:velocity_width}
\end{figure*}

\section{Discussion}\label{sec:discussion}
We derived the velocity widths in the beamsize both for simulation and observations.
The velocity widths used here are
FWZI and
intensity-weighted velocity dispersion ($\sigma_{\rm disp}$).
Simple Gaussian fit would not be  suitable
for the estimation of the velocity widths  
because the profiles in figure \ref{fig:fujimoto} and most of the observed profiles in figure \ref{fig:result}
show asymmetric shapes and
some of them show a double peak emission.

The $\sigma_{\rm disp}$ is defined as the second moment of a spectrum,
\begin{equation}
\sigma_{\rm disp} = \left[ \frac{\Sigma (v_i - \braket{v})^2 T_{\rm mb}(v_i)}{\Sigma T_{\rm mb}(v_i)} \right]^{1/2}
\end{equation}
where $i$ represents each velocity bin and $\braket{v}$ is the intensity-weighted mean velocity (the first moment),
\begin{equation}
\braket{v} = \frac{\Sigma v_i T_{\rm mb}(v_i)}{\Sigma T_{\rm mb}(v_i)}.
\end{equation}

Firstly, we derived the velocity widths of the simulated line profiles shown in figure \ref{fig:fujimoto}.
In the bar regions, FWZI and $\sigma_{\rm disp}$ of black (red) line  are
210 (270) $\rm km~s^{-1}$ and 38 (55) $\rm km~s^{-1}$, respectively.
In the arm regions, FWZI and $\sigma_{\rm disp}$ of blue (green) line  are
120 (150) $\rm km~s^{-1}$ and 20 (29) $\rm km~s^{-1}$, respectively.
The velocity widths in the bar regions are significantly larger than
those in the arm regions in the simulation.

Next, we derived the velocity widths of the observed spectra.
The FWZI is taken as  the width of the hatched region  in figure \ref{fig:result}.
The typical uncertainty of resulting FWZI in figure \ref{fig:result} is estimated to be $20 \sim 30~\rm km~s^{-1}$.
The error of $\sigma_{\rm disp}$ is calculated by considering the error propagation of equation (3) and $T_{\rm rms}$ in table \ref{tab:obsresult}
and is typically $10~\rm km~s^{-1}$.
The resulting velocity widths are summarized in table \ref{tab:linewidth} and shown in figure \ref{fig:velocity_width}.
In NGC 1300, the mean of FWZI and $\sigma_{\rm disp}$ of the bar (arm) regions  are
120 (83) $\rm km~s^{-1}$ and 28 (17) $\rm km~s^{-1}$, respectively.
In NGC 5383, the mean of FWZI and $\sigma_{\rm disp}$ of the bar (arm) regions  are
190 (130) $\rm km~s^{-1}$ and 42 (30) $\rm km~s^{-1}$, respectively.
Thus, FWZI and $\sigma_{\rm disp}$ in the bar regions are larger than those in the arm regions on average in each galaxy,
which is consistent with the new scenario.
Compared with the individual regions in NGC 1300, however, 
the velocity width in Bar B region is comparable to that in Arm B region, and
in the Arm C region, where no clear HII region is seen,
the velocity width is smaller than those in the Arm A, B, and the bar regions.
In NGC 5383, the velocity width in Bar 2 is comparable to those in the arm regions.

The velocity widths in NGC 1300 and NGC 5383 show
the qualitatively similar tendency seen in the simulation (figure \ref{fig:fujimoto});
the velocity widths in the bar regions are larger than those in the arm regions on average,
although there are exceptions among individual regions.
Similar tendencies are seen in the previous observations toward other intermediate-type (SAB) barred galaxies
\citep{Regan99,Sorai12,Morokuma15,Muraoka16}.

The velocity widths in NGC 5383 are systematically larger than those in NGC 1300,
and the velocity widths in the bar regions of NGC 1300 are
comparable to those in the arm regions of NGC 5383.
Such a simple comparison of the velocity widths is, however, not appropriate.
The velocity widths in the beamzsize are 
affected by the velocity field,
the distribution of molecular clouds,
and their relative velocity among the molecular clouds ($\sigma_{\rm gas}$) in the beamsize.
Since the beamsize is larger in NGC 5383 than that in NGC 1300
in physical scale,
there are possibilities that 
the larger velocity range is covered
and/or wider velocity components of the molecular clouds reside in the beamsize of NGC 5383.
The molecular gas mass in the beamsize is larger in NGC 5383 than
that in NGC 1300.
This may also be related to the difference of the velocity width.

To derive $\sigma_{\rm gas}$ in the beam,
we need to know the gas velocity field and distribution of molecular clouds in the beam.
However, it is very difficult to achieve them with a single-dish telescope.
Therefore, although our results are broadly consistent with the new scenario,
concluding that the results support the new scenario is premature.
Observations with ALMA/NOEMA
are indispensable to examine the scenario.

\section{Summary} \label{sec:summary}
We made $^{12}$CO($J = 1 -0$) observations toward bar and arm regions of the strongly barred galaxies,
NGC 1300 and NGC 5383 with the Nobeyama 45-m telescope.
We detected CO emissions from all the regions,
and this indicates that the molecular gases do exist in the
strong bars with no clear HII regions.
In both galaxies,
the velocity width tends to be larger in the bar region than in the arm region,
which is qualitatively consistent with the new idea 
for the suppression of the star formation in the bar region,
i.e., the high speed collisions of the molecular clouds in the bar region suppress the massive star formation.
However, the trend is not so clear.
Because the velocity width is affected by the velocity field,
molecular cloud distribution,
and their relative velocity in the beam,
further observations of higher angular resolution with ALMA/NOEMA are necessary.

\begin{table}
  \tbl{Velocity widths}{%
  \begin{tabular}{clcl}
  \hline
  \multicolumn{2}{c}{Region}      & \multicolumn{1}{c}{FWZI$^{\ast)}$ }   & \multicolumn{1}{c}{$\sigma_{\rm disp}$}\\
                          &             & \multicolumn{1}{c}{($\rm km~s^{-1}$) }& \multicolumn{1}{c}{($\rm km~s^{-1}$)}  \\
  \hline
  \hline
Simulation                & Bar (black) & $210$    & $42$ \\
(Fig. \ref{fig:fujimoto}) & Bar (red)   & $270$    & $56$ \\
                          & Arm (blue)  & $120$    & $21$ \\
                          & Arm (green) & $150$    & $30$ \\
  \hline
NGC 1300                  & Bar A       & $130$    & $33 \pm 9$ \\
(Fig. \ref{fig:result})   & Bar B       & $110$    & $22 \pm 7$ \\
                          & Arm A       & $\ \ 80$ & $17 \pm 9$ \\
                          & Arm B       & $100$    & $22 \pm 12$ \\
                          & Arm C       & $\ \ 70$ & $12 \pm 10$ \\
                          & Bar End A   & $\ \ 90$ & $19 \pm 10$ \\
\hline
NGC 5383                  & Bar 1       & $240$    & $52 \pm 11$ \\
(Fig. \ref{fig:result})   & Bar 2       & $140$    & $32 \pm 11$ \\
                          & Arm 1       & $150$    & $34 \pm 8$ \\
                          & Arm 2       & $110$    & $25 \pm 6$ \\
                          & Bar End 1   & $170$    & $46 \pm 13$ \\
  \hline
    \end{tabular}}\label{tab:linewidth}
\begin{tabnote}
$^{\ast)}$ Typical uncertainty of resulting FWZI in figure \ref{fig:result} is estimated to be $20 \sim 30~\rm km~s^{-1}$.
\end{tabnote}
\end{table}

\begin{ack}
We acknowledge the members of Nobeyama Radio Observatory for their help during the observations
and A. Seko for his  useful comments and support of the observations.
We also thank K. Sorai for useful discussion and 
K. Sheth for making his data available. 
We further want to thank the referee for comments.
K.O. is supported by the Grant-in-Aid for Scientific Research (C) (16K05294) from the Japan Society of the Promotion of Science (JSPS). 
\end{ack}


\end{document}